\documentclass{ws-procs9x6}

\begin{document}

\title{New H1 Results on Isolated Leptons and Missing $P_{T}$ at HERA}

\author{D.~M.~SOUTH\footnote{\uppercase{O}n behalf of the \uppercase{H}1
\uppercase{C}ollaboration.}}

\address{Deutsches Elektronen Synchrotron\\
Notkestrasse 85, 22607, Hamburg, Germany\\
E-mail: David.South@desy.de }

\maketitle

\abstracts{The search for events containing isolated leptons
(electrons or muons) and missing transverse momentum produced in $e^\pm p$
collisions is performed with the H1 detector at HERA in the period 1994--2005.
The analysed data sample corresponds to an integrated luminosity of
279~pb$^{-1}$, which includes 53~pb$^{-1}$ of $e^{+}p$ data and 107~pb$^{-1}$
of $e^{-}p$ data from the new HERA~II phase. A total of 40 events
are observed in the data, compared to a Standard Model (SM) prediction of
34.3~$\pm$~4.8. At large hadronic transverse momentum $P_{T}^{X} >$~25~GeV,
a total of 17 events are observed compared to 9.0~$\pm$~1.5 predicted by
the SM. In this region, 15 events are observed in the $e^{+}p$ data
compared to a SM prediction of 4.6~$\pm$~0.8, whereas in the $e^{-}p$ data
2 events are observed compared to a SM prediction of 4.4~$\pm$~0.7.}

\section{Introduction}

Events containing a high $P_{T}$ isolated electron or muon and associated
with missing transverse momentum have been observed at
HERA\cite{isoleph1origwpaper,isoleph1newwpaper,isolepzeusorigwpaper,zeustop}.
An excess of HERA~I (1994--2000) data events compared to the SM prediction at large
hadronic transverse momentum $P_{T}^{X}$ was reported by the H1
Collaboration\cite{isoleph1newwpaper}, which was not confirmed by the ZEUS
Collaboration, although using a slightly different analysis approach\cite{zeustop}.
Most of the HERA~I data, luminosity 118~pb$^{-1}$, were taken in $e^{+}p$ collisions. 
The H1 analysis has been updated\cite{isoleph1hera2,isoleph1hera2new} to include
new $e^{\pm}p$ data from the ongoing HERA~II phase (2003--2005), resulting in a
total analysed luminosity of 279~pb$^{-1}$.

\section{Standard Model Signal Processes}

\begin{figure}[t] 
  \begin{flushleft}
      \includegraphics[width=.60\textwidth]{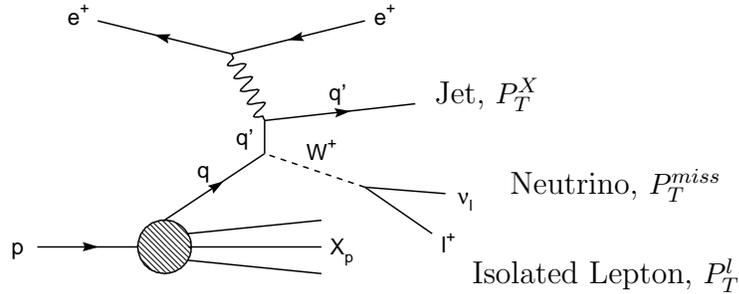}
  \end{flushleft}
\begin{picture} (0.,0.)
\setlength{\unitlength}{1.0cm}
\put (6.0,3.4){\large Jet, $P_{T}^{X}$} 
\put (7.0,2.2){\large Neutrino, $P_{T}^{miss}$} 
\put (6.5,1.0){\large Isolated Lepton, $P_{T}^{l}$} 
\end{picture}
\vspace*{-20pt}
\caption{Feynman diagram of the process $ep \rightarrow eW^{\pm}$($\rightarrow l\nu$)$X$,
	which is the main SM contribution to the search for events with isolated leptons
	and missing transverse momentum. The main final state components are also labelled.}
\label{fig:feynmandiagram}
\end{figure}

The signal topology in this analysis is a prominent, isolated lepton accompanied by large,
genuine missing transverse momentum. The main SM contribution to such a topology comes
from the production of real $W$ bosons with subsequent leptonic decay
$ep \rightarrow eW^{\pm}$($\rightarrow l\nu$)$X$, as illustrated in figure
\ref{fig:feynmandiagram}. The struck quark quickly hadronises giving rise to the
(typically low $P_{T}$) hadronic system $X$, whilst the $W$ decay neutrino escapes
undetected, resulting in a substantial transverse momentum imbalance in the event,
$P_{T}^{miss}$. Additional, smaller signal contributions arise from the production of
$W$ bosons via the equivalent charged current process
$ep \rightarrow \nu$$W^{\pm}$($\rightarrow l\nu$)$X$ and the production of $Z^{0}$ bosons
with subsequent decay to neutrinos $ep \rightarrow eZ^{0}$($\rightarrow \nu\bar{\nu}$)$X$,
which contributes only to the electron channel.

\section{Event Selection}

The event selection employed is identical to that used in the HERA~I
analysis\cite{isoleph1newwpaper}. The kinematic phase space is defined
as follows: The identified lepton should have high transverse momentum
$P_{T}^{l} >$~10~GeV, be observed in the central region of the detector
5$^{\circ}$~$< \theta_{l} <$~140$^{\circ}$ and be isolated with respect
to jets and other tracks in the event. The event should also contain a large
transverse momentum imbalance, $P_{T}^{miss} >$~12~GeV. Further cuts are then applied,
which are designed to reduce SM background, whilst preserving a high level of
signal purity. Event quantities sensitive to the presence of high energy undetected
particles in the event are employed such as the azimuthal balance of the event, the
difference in azimuthal angle between the lepton and the hadronic system and the
longitudinal momentum imbalance. To ensure that the two lepton channels are
exclusive and may therefore be combined, electron events must contain no
isolated muons.

\section{Results}

\begin{figure}[t]
  \centering
  \includegraphics[width=.49\textwidth]{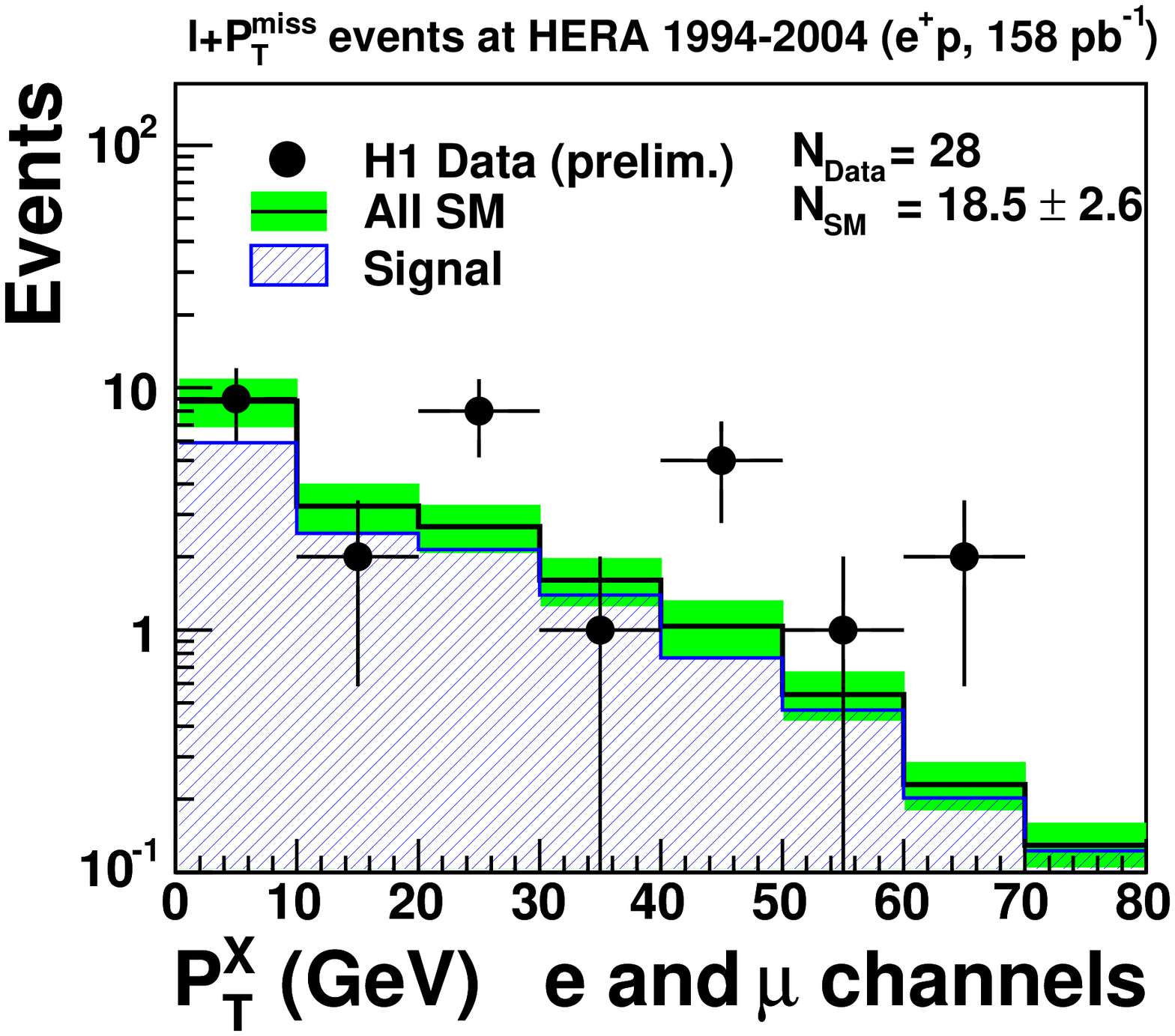}
  \includegraphics[width=.49\textwidth]{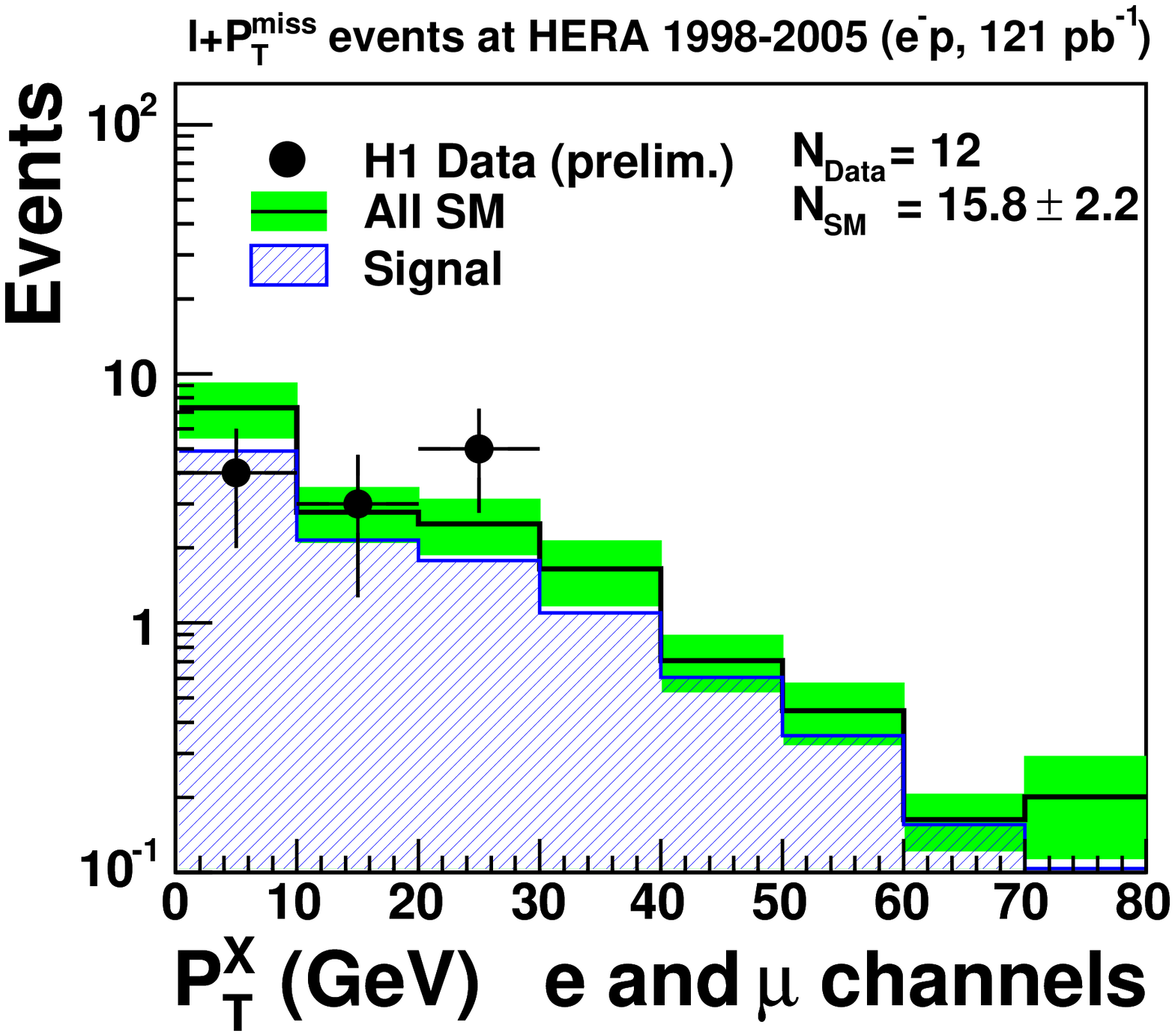}
  \caption{The hadronic transverse momentum spectra of the observed events in the H1
	isolated lepton analysis. The $e^{+}p$ data sample is shown on the left and
	the $e^{-}p$ data sample on the right. The data are the points, the full histogram
	is the SM expectation and the shaded band is the total SM error. The signal
	component, dominated by real $W$ production, is shown by the hatched histogram.}
\label{fig:spectra}
\end{figure}

\begin{figure}[h] 
  \centering	
  \includegraphics[width=.98\textwidth]{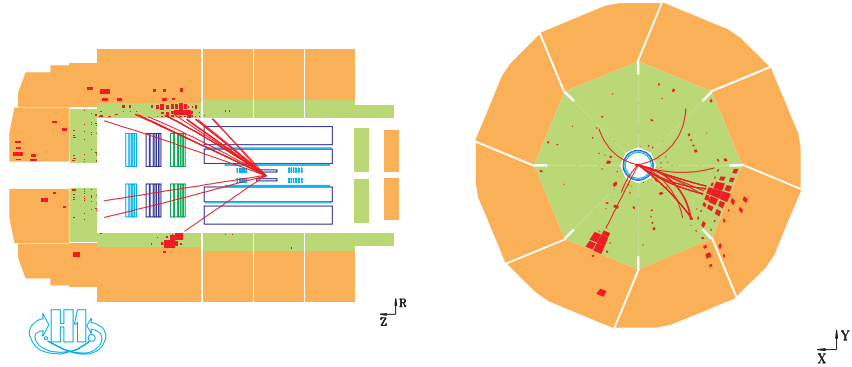}
  \caption{Display of an event with an isolated electron, missing transverse momentum and a
	prominent hadronic jet recorded by the H1 experiment in the HERA~II $e^{+}p$ data.}
\label{fig:highptxevent}
\end{figure}

In the final event sample a total of 40 events are observed in the H1 data, compared
to a SM prediction of 34.3~$\pm$~4.8. The hadronic transverse momentum spectra of
the $e^{\pm}p$ data are presented in figure \ref{fig:spectra}. At large values of
$P_{T}^{X}$, a kinematic region atypical of SM $W$ production, an excess of $e^{+}p$
data events is observed over the SM expectation, as can be seen in figure
\ref{fig:spectra} (left). For $P_{T}^{X} >$~25~GeV a total of 15 data events are
observed in the $e^{+}p$ data compared to a SM prediction of 4.6~$\pm$~0.8, equivalent
to a fluctuation of approximately 3.4$\sigma$. Figure \ref{fig:highptxevent} shows an
event in the HERA~II $e^{+}p$ data containing an isolated electron, missing $P_{T}$ and
a hadronic jet with large $P_{T}^{X}$. Interestingly, a similar excess is not observed
in the current $e^{-}p$ data sample, as can be seen in figure \ref{fig:spectra} (right),
where 2 data events are observed compared to a SM prediction of 4.4~$\pm$~0.7. The
$e^{-}p$ data sample now includes almost a factor of 10 increase in statistics with
respect to the HERA~I data set. A summary of the results is presented
in table \ref{tab:isolep}.

\section{Summary}

The search for events containing high $P_{T}$ isolated electrons or muons and
missing transverse momentum produced in $e^\pm p$ collisions is performed using data
collected by the H1 detector at HERA in the period 1994--2005, corresponding to an
integrated luminosity 279~pb$^{-1}$. At large values of $P_{T}^{X} >$~25~GeV an excess
of events is observed in the $e^{+}p$ data sample, where 15 events are observed compared
to a SM prediction of 4.6~$\pm$~0.8. No such excess is observed in the $e^{-}p$ data
sample or in the recent re-analysis performed by the ZEUS
Collaboration\cite{isolepzeushera2,zeustheseproceedings}. The continued increase in
luminosity from the HERA~II programme will hopefully clarify the observed H1 excess
in the $e^{+}p$ data at large hadronic transverse momentum.

\vspace*{-10pt}

\begin{table}[h]
\tbl{Summary of the H1 search for events with isolated electrons or muons and missing
	transverse momentum, in the kinematic region $P_{T}^{X}>$~25~GeV. The number
	of observed events is compared to the SM prediction for the $e^{+}p$,
	$e^{-}p$ and $e^{\pm}p$ data sets.}
{\small
\begin{tabular}{|c||c|c|c|}
\hline
H1 Preliminary & e channel & $\mu$ channel & combined e \& $\mu$\\
$P_{T}^{X} >$ 25 GeV & obs./exp. & obs./exp. & obs./exp. \\
\hline
\hline
1994--2004 $e^{+}p$ 158 pb$^{-1}$ &
9 / 2.3 $\pm$ 0.4 & 6 / 2.3 $\pm$ 0.4 & 15 / 4.6 $\pm$ 0.8 \\
\hline
1998--2005 $e^{-}p$ 121 pb$^{-1}$ &
2 / 2.4 $\pm$ 0.5 & 0 / 2.0 $\pm$ 0.3 &  2 / 4.4 $\pm$ 0.7 \\
\hline
\hline
1994--2005 $e^{\pm}p$ 279 pb$^{-1}$ &
11 / 4.7 $\pm$ 0.9 & 6 / 4.3 $\pm$ 0.7 & 17 / 9.0 $\pm$ 1.5 \\
\hline
\end{tabular}
\label{tab:isolep}}
\vspace*{-10pt}
\end{table}

\end{document}